\documentclass[onecolumn]{aastex631}

\pdfoutput=1

\shorttitle{Juno ICME Catalog}
\shortauthors{E.E.Davies et al.}
\graphicspath{{./}{figures/}}

\begin{document}

\title{A Catalog of Interplanetary Coronal Mass Ejections Observed by Juno between 1 and 5.4~AU}

\author[0000-0001-9992-8471]{Emma E. Davies}
\affiliation{Institute for the Study of Earth, Ocean, and Space, University of New Hampshire, Durham, New Hampshire, USA}
\affiliation{Department of Physics, Imperial College London, London, UK}
\correspondingauthor{Emma E. Davies}
\email{emma.davies@unh.edu}

\author[0000-0003-2701-0375]{Robert J. Forsyth}
\affiliation{Department of Physics, Imperial College London, London, UK}

\author[0000-0002-9276-9487]{Réka M. Winslow}
\affiliation{Institute for the Study of Earth, Ocean, and Space, University of New Hampshire, Durham, New Hampshire, USA}

\author[0000-0001-6868-4152]{Christian M\"ostl}
\affiliation{Space Research Institute, Austrian Academy of Sciences, Graz, Austria}

\author[0000-0002-1890-6156]{Noé Lugaz}
\affiliation{Institute for the Study of Earth, Ocean, and Space, University of New Hampshire, Durham, New Hampshire, USA}

\begin{abstract}

We use magnetic field measurements by the Juno spacecraft to catalog and investigate interplanetary coronal mass ejections (ICMEs) beyond 1~AU. During its cruise phase, Juno spent about 5 years in the solar wind between 2011 September and 2016 June, providing measurements of the interplanetary magnetic field (IMF) between 1 and 5.4~AU. Juno therefore presents the most recent opportunity for a statistical analysis of ICME properties beyond 1~AU since the Ulysses mission (1990--2009). Our catalog includes 80 such ICME events, 32 of which contain associated flux-rope-like structures. We find that the dependency of the mean magnetic field strength of the magnetic flux ropes decreases with heliocentric distance as $r^{-1.24 \pm 0.43}$ between 1 and 5.4~AU, in good agreement with previous relationships calculated using ICME catalogs at Ulysses. We combine the Juno catalog with the HELCATS catalog to create a dataset of ICMEs covering 0.3--5.4~AU. Using a linear regression model to fit the combined dataset on a double-logarithmic plot, we find that there is a clear difference between global expansion rates for ICMEs observed at lesser heliocentric distances and those observed farther out beyond 1~AU. The cataloged ICMEs at Juno present a good basis for future multispacecraft studies of ICME evolution between the inner heliosphere, 1~AU, and beyond. 

\end{abstract}

\keywords{Solar coronal mass ejections(310) --- Heliosphere(711) --- Dynamical evolution(421)}

\section{Introduction} \label{sec:intro}

Interplanetary coronal mass ejections (ICMEs) are large-scale structures of plasma and magnetic field that are driven from the solar atmosphere and propagate through the heliosphere. These transient structures are distinguished from the ambient solar wind in situ by distinct magnetic field and plasma signatures amongst other composition and charge-state features \citep{zurbuchen2006situ}. ICMEs have long been causally linked with geomagnetic activity at Earth \citep{gosling1991geo, kahler1992solar} and are the main drivers of severe space weather at Earth \citep[e.g.][]{eastwood2008science, kilpua2017geoeffective}. 

ICMEs are the interplanetary counterparts of coronal mass ejections (CMEs). In situ measurements of ICMEs often display signatures of magnetic flux rope structures \citep{cane2003interplanetary}, likely the interplanetary manifestations of those observed remotely in coronagraph images \citep{burlaga1982magnetic}. Such ICMEs are known as magnetic clouds, defined as having features that include an enhanced magnetic field, smooth rotation of the magnetic field vector, low plasma $\beta$ and a low proton temperature \citep{burlaga1981magnetic}. Magnetic clouds typically exhibit well-structured magnetic fields consistent with force-free flux ropes \citep{goldstein1983field}.

As ICMEs propagate through the heliosphere, many processes can affect their evolution. These include radial deflections that can change the course of an ICME significantly \citep{gosling1987deflect}, rotation or changes to the inclination of the flux rope \citep[e.g.][]{nieves2013inner}, and erosion of the magnetic flux rope via magnetic reconnection between the ICME and the magnetic field of the solar wind \citep{mccomas1998solar, dasso2006new, ruffenach2012multispacecraft, ruffenach2015statistical}. ICME evolution can also be complicated by interactions between ICMEs \citep[e.g.][]{lugaz2017importance, wang2018understanding} and with other large-scale features such as the heliospheric current sheet (HCS), stream interaction regions (SIRs), and high-speed streams (HSSs) \citep[e.g.][]{winslow2016longitudinal,davies2020radial,winslow2021first,winslow2021effect}, and even by the solar wind environment through which they propagate, kinematically distorting the large-scale structure of the ICME \citep[e.g.][]{owens2006magnetic, savani2010observational, owens2017coronal, davies2021}. The geoeffectiveness of an ICME depends strongly on the magnitude and the orientation of the magnetic field within the ICME \citep[e.g.][]{bothmer1998structure, huttunen2005properties} and therefore understanding ICME evolution and interaction with the solar wind and other structures is of great interest in space weather forecasting.

ICMEs have been studied in situ at varying heliocentric distances, using dedicated solar wind spacecraft such as Helios \citep[0.3--1~AU, e.g.][]{cane1997helios, bothmer1998structure}, ACE/Wind, STEREO \citep[at 1~AU, e.g.][]{cane2003interplanetary, richardson2010near, jian2018stereo}, and Ulysses \citep[1--5.4~AU, e.g.][]{liu2005statistical, wang2005characteristics, ebert2009bulk, du2010interplanetary, richardson2014identification}, based on a variety of solar wind signatures such as enhancements of the magnetic field, occurrence of abnormally low proton temperatures, and galactic cosmic ray (GCR) decreases as primary ICME identifiers. Other studies of ICMEs in the inner heliosphere include the use of planetary mission spacecraft such as MESSENGER and Venus Express \citep{winslow2015interplanetary, good2016interplanetary}, where ICMEs were identified by only their magnetic field signatures. 

Catalogs such as these are crucial in understanding the in situ properties of ICMEs at different heliocentric distances, and combining different catalogs can shed light on ICME evolution in the heliosphere. \citet{janvier2019generic} compared the average properties of ICMEs at Mercury, Venus, and Earth, using the catalogs of \citet{winslow2015interplanetary}, \citet{good2016interplanetary} and \citet{richardson2010near}, respectively. The magnetic field profiles of ICMEs were found to change with heliocentric distance for different velocity categories of ICMEs: generally, the magnetic field profiles were found to be more symmetric for slower ICMEs, in contrast to faster ICMEs with more asymmetric profiles, where the magnetic field magnitude was larger at the front of the magnetic flux rope than towards the trailing edge. The ICME profiles identified at Mercury were found to be more asymmetric than those at Earth, implying a relaxation of the magnetic field profile as the ICME propagates.

To better understand ICME evolution in situ, it is also useful to track signatures of specific ICMEs over large heliocentric distances. \citet{salman2020radial} utilized the catalogs of \citet{winslow2015interplanetary}, \citet{good2016interplanetary}, \citet{richardson2010near} and \citet{jian2018stereo} to study the variation of properties (e.g. propagation speed, acceleration, magnetic field strength) of 47 ICMEs observed in longitudinal conjunction by at least two spacecraft as they propagated through the inner heliosphere. The global expansion of the ICMEs with increasing heliocentric distance was found to be consistent with previous statistical trends, but individual events displayed significant variability when compared to average trends. The same ICMEs observed in conjunction across the catalogs listed were used by \citet{lugaz2020inconsistencies} to investigate the relationship between the global and local expansion of ICMEs. The two measures of expansion were found to have little relation with each other; however, the expansion was found to depend on the initial magnetic field strength within the ICME ejecta.

Studies of ICME evolution beyond 1~AU are even less common than those $<$1~AU. The Ulysses mission \citep[e.g.][]{wenzel1992ulysses} provided comprehensive measurements of the heliosphere and solar wind beyond 1~AU and at high latitudes following its launch in 1990 October. \citet{ebert2009bulk} identified 178 ICMEs in the Ulysses data between 1992 and 2007 using a variety of magnetic and plasma properties. Most ICMEs were identified at latitudes less than 40$^\circ$, and only small latitudinal variations of ICME properties were observed. \citet{du2010interplanetary} identified 181 ICMEs at Ulysses using low proton temperature as the primary identifier between 1991 and 2007 (the study is an extension of \citet{wang2005characteristics}). Forty-three percent of ICMEs identified could be classified as magnetic clouds. The study found that the occurrence rate of ICMEs approximately follows the solar activity cycle, further confirming the findings of \citet{wang2005characteristics}, and in agreement with the \citet{richardson2010near} study of ICMEs at 1~AU. \citet{richardson2014identification} identified 279 ICMEs over the Ulysses mission duration (1990--2008), and found that the latitudinal distribution of ICMEs is related to the level of solar activity; around solar minimum, most ICMEs were identified at latitudes below 50$^{\circ}$, whereas around solar maximum, ICMEs were observed by Ulysses at latitudes up to the 80$^{\circ}$ reached by the spacecraft. \citet{richardson2014identification} also provided a comparison of the number of ICMEs identified in their study with those of \citet{ebert2009bulk} and \citet{du2010interplanetary} and found that only 102 ICMEs were identified across all three catalogs (relative to 327 in total), demonstrating the highly subjective nature of ICME identification and dependence on identification criteria. 

Similarly to the ICME catalogs of \citet{winslow2015interplanetary} and \citet{good2016interplanetary}, this study makes use of the magnetic field data taken by the planetary mission spacecraft, Juno \citep{bolton2010juno, bolton2017juno} during its cruise phase to Jupiter. Juno spent $\sim 5$ years in the solar wind between 2011 September and 2016 June, providing measurements of the solar wind between 1 and 5.4~AU. The Juno spacecraft therefore presents the most recent opportunity for a statistical analysis of ICME properties beyond 1~AU since the Ulysses mission (1990--2009).

In this study we present a catalog of ICMEs identified at Juno and a statistical analysis of their properties between 1 and 5.4~AU. Section \ref{sec:junodataoverview} presents an overview of the Juno mission and the availability of magnetometer data throughout its cruise phase, and Section \ref{sec:identification_criteria} details the identification criteria used to compile the ICME catalog. Section \ref{sec:juno_catalog} presents an analysis of the observed ICME properties in this catalog and a comparison with dependencies previously found in catalogs covering similar heliocentric distance ranges. Section \ref{sec:juno_catalog} also includes ICMEs identified by spacecraft at heliocentric distances of 1~AU and below to provide a statistical picture of ICME evolution between 0.3 and 5.4~AU. A summary of findings is presented in Section \ref{sec:summary}. 

\section{Overview of Juno Cruise Data} \label{sec:junodataoverview}

\begin{figure}[t]
\centering
\includegraphics[width=\linewidth]{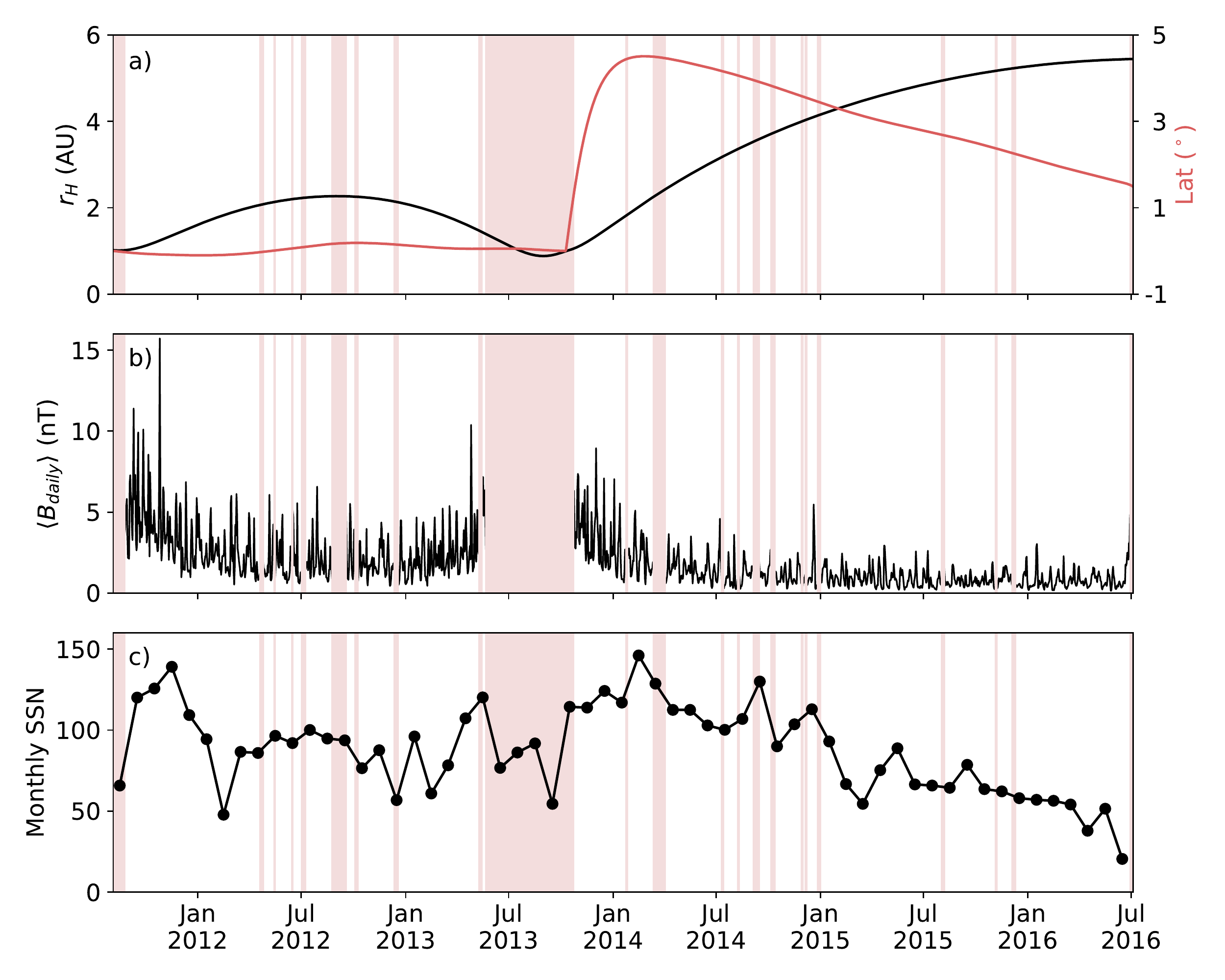}
\caption{Overview of Juno cruise phase data. (a) Heliocentric distance (left) and latitude (right) of the Juno orbit in heliocentric ecliptic coordinates (HAE/HEE) between launch on 2011 August 5 and orbital insertion at Jupiter on 2016 July 4. (b) The mean daily magnetic field magnitude observed by Juno during its cruise phase. Vertical lines indicate days for which magnetometer data are unavailable. (c) Monthly mean sunspot number.}
\label{fig:dataoverview}
\end{figure}

The magnetic field measurements used in this study were taken by the Juno  Magnetic  Field  Investigation instrument \citep[MAG;][]{connerney2017juno} and are available via the Planetary Plasma Interactions (PPI) node of the Planetary Data System (PDS) archive at \url{https://pds-ppi.igpp.ucla.edu/}. Data is available at both 1 minute and 1 second cadence, in four coordinate systems: planetocentric (PC), payload (PL), solar equatorial (SE), and Sun state (SS). For this study, we use data at a 1 minute time resolution and in SE coordinates, equivalent to radial-tangential-normal (RTN) coordinates. Data from other instruments are not available during the cruise phase, with the exception of the Jovian Auroral Distributions Experiment \citep[JADE;][]{mccomas2017jovian}, for which plasma data are available during the final month of the cruise phase. 

Figure \ref{fig:dataoverview} presents an overview of (a) the spacecraft heliocentric distance and latitude between launch on 2011 August 5 and orbital insertion at Jupiter on 2016 July 4, (b) the magnetic field data and its availability, and (c) the monthly mean sunspot number (\url{http://www.sidc.be/silso/}) recorded throughout the cruise phase. Figure \ref{fig:dataoverview}a shows that following launch, Juno moved away from the Sun with an aphelion of $\sim$2~AU whilst remaining close to the ecliptic plane before a gravity assist at Earth redirected its trajectory towards Jupiter at 5.4~AU. The gravity assist at Earth resulted in raising the heliolatitude of the spacecraft to $\sim4.5^{\circ}$ which gradually declined back towards the ecliptic plane with increasing heliocentric distance. The largest data gap during the cruise phase occurred during the months surrounding the gravity assist at Earth, between 2013 May 23 and October 5 inclusive. However, despite data gaps, Figure \ref{fig:dataoverview} shows that, overall, there is good data availability over the cruise phase for the complete range of heliocentric distances covered.

Figure \ref{fig:dataoverview}b presents the mean daily  magnetic field magnitude observed by Juno during the cruise phase. The magnetic field magnitude is inversely proportional to the heliocentric distance, $r$, well approximated by $B \propto \sqrt{(1 + r^2)}/r^2$ which approaches $B \propto 1/r$ with increasing heliocentric distance \citep[e.g.][]{parker1958dynamics, burlaga1984large}. \citet{gruesbeck2017interplanetary} found that applying the expected $1/r$ relationship to the Juno cruise data resulted in a very good fit to the magnetic field data. The magnetic field strength throughout the cruise phase was also found to be of lower magnitude in comparison to observations of previous solar cycles made by Voyager and Ulysses. This is consistent with observations of the radial magnetic field at 1 AU during Solar Cycle 24, where the magnitude was found to be $\sim$38\% lower than previous solar cycles observed between the mid-1970s to the mid-1990s \citep{mccomas2013weakest} and may have implications on the expansion of ICMEs.

The monthly mean sunspot number presented in Figure \ref{fig:dataoverview}c shows that Juno was launched as the solar activity of Solar Cycle 24 was increasing, with the first peak in sunspot number occurring in 2011 November only a few months later. Most of the cruise phase took place under solar maximum conditions with activity peaking in 2014 April and slowly declining thereafter. It is therefore not possible for this study to provide a comparison of ICME occurrence rates and other parameters over the different stages of the solar cycle; however, this timing presents ideal conditions for observing ICMEs. 

\section{ICME Identification Criteria} \label{sec:identification_criteria}

ICMEs can be distinguished from the ambient solar wind by distinct features identified in the magnetic field and plasma data (as well as other compositional and charge-state features) measured in situ \citep[e.g.][]{zurbuchen2006situ}. With the absence of plasma data, the identification of potential ICME candidates from magnetic field data alone is susceptible to uncertainty, and therefore, we focus on identifying features in the magnetic field data that meet three strict criteria: 1) an enhanced magnetic field magnitude at least twice that of the expected ambient interplanetary magnetic field (IMF); 2) a magnetic field profile typical of an ICME, e.g., a shock-like discontinuity (defined as a step-function-like increase in the magnetic field strength), sheath, and region of magnetic ejecta with decreased magnetic field variance; and 3) a duration on the order of at least one day. Identification of ICMEs at Juno has been complemented wherever possible by magnetic field and plasma observations taken by other spacecraft (typically located at 1~AU) when in close conjunction with Juno and by solar wind models such as the WSA-ENLIL-cone model \citep{odstrcil2003enlil}. 

As the heliocentric distance at which Juno observes events is continually changing, the expected magnetic field strength of the ambient IMF with increasing heliocentric distance has been taken into account by fitting a $1/r$ relationship to the magnetic field data \citep[see][]{gruesbeck2017interplanetary}. Days within which the magnetic field magnitude exceeds twice the estimated value for nominal conditions at the heliocentric distance at which they are observed have been automatically selected as enhanced. This selection criterion is consistent with other studies such as the \citet{klein1982interplanetary} study of ICMEs at 1~AU, where selected events meet the enhancement criterion $>$10~nT, approximately twice that of the background IMF at 1~AU which is on the order of $\sim$~5~nT.

Days in which the magnetic field magnitude met the enhancement criterion have been grouped together and inspected for features typical of ICME magnetic field profiles. Almost all enhancement groupings identified in the dataset (including both potential ICME candidates and other large-scale structures) were associated with a forward-shock-like discontinuity driven by the enhancement. Previous studies have found that 51\% of ICMEs identified near Earth drove upstream shocks \citep[][]{richardson2010near}, and similarly, 58\% of ICMEs at 5.3~AU \citep[][]{jian2008stream}. We therefore recognize that by identifying shock-driving ICMEs, the catalog is biased towards fast ICMEs with strong magnetic fields and potentially excludes up to half of the ICMEs that may be present in the dataset. However, by excluding those without a shock-like discontinuity, we ensure that all events identified in the Juno dataset are likely ICMEs and avoid false-positive identifications. 

To distinguish potential ICME candidates from other enhanced features such as stream interaction regions (SIRs) or corotating interaction regions (CIRs), periods of lower-variance magnetic field associated with magnetic ejecta of ICMEs were identified by visual inspection. We also observed the overall magnetic field profile of the enhancement: although not an exclusive feature, forward/reverse shock pairs are often associated with SIR/CIRs beyond 1~AU \citep{richardson2004cirs,jian2008stream}. The end of an ICME can be indicated by either a reverse shock, a discontinuity in the magnetic field components, or the magnetic field magnitude gradually returning to ambient IMF values.

\begin{figure*}
\gridline{\fig{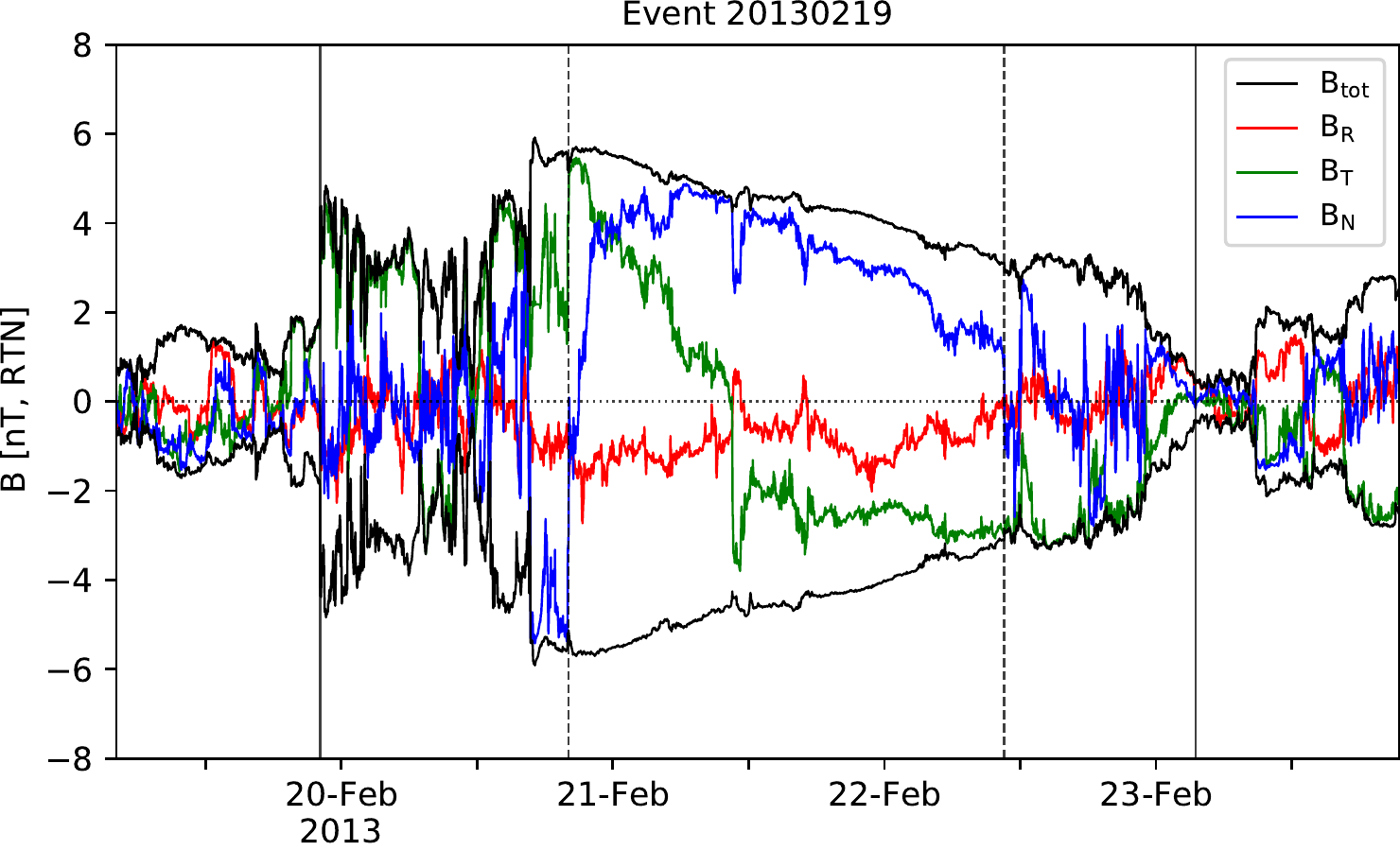}{0.45\textwidth}{(a)}
\fig{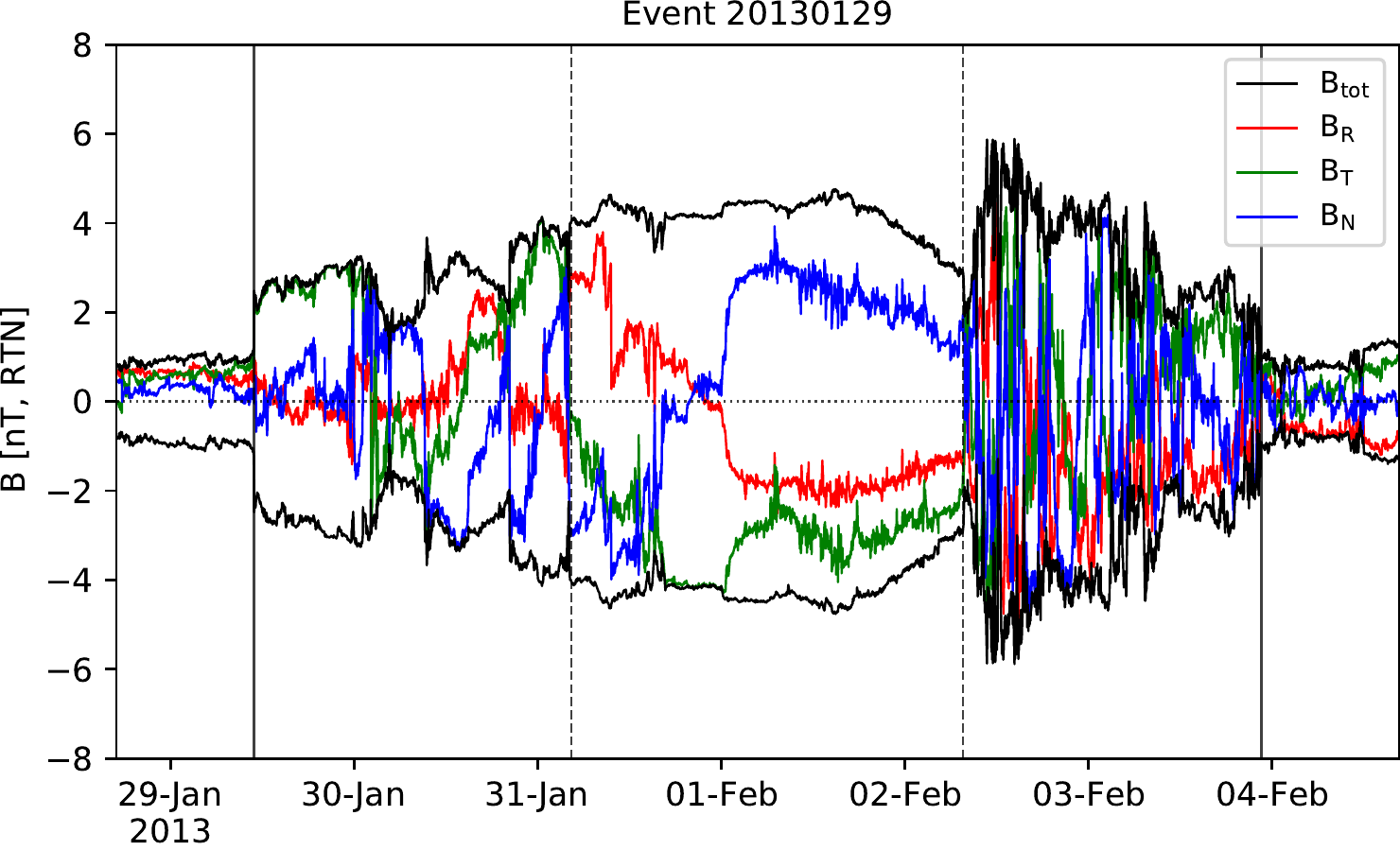}{0.45\textwidth}{(b)}
}
\caption{Examples of events identified in the Juno catalog during a similar time period, whilst the spacecraft was located at a heliocentric distance of $\sim$~2~AU. Solid vertical lines mark the start and end of the ICME, and the dashed lines delineate the magnetic flux rope. (a) Event 20130219 has been classified as an ICME due to features in the magnetic data, including a shock-like discontinuity, compressed sheath region, and low-variance region of ejecta comprising a magnetic flux rope structure. (b) Event 20130129 provides an example of a merged interaction region (MIR) due to the compressed solar wind material of higher magnitude following the flux-rope-like structure.}
\label{fig:exampleplots}
\end{figure*}

Many events cataloged therefore exhibit magnetic field profiles typical of ICMEs including a shock-like discontinuity, followed by a sheath and low-variance periods of an enhanced magnetic field. Of these candidates, some also include a relatively smooth rotation of the magnetic field components associated with a magnetic-flux-rope-like structure, a feature of the subset of ICMEs known as magnetic clouds \citep{burlaga1981magnetic}. Without plasma data, it is not possible to confirm whether these events also exhibit the low proton temperature and plasma $\beta$ necessary to be classified as a magnetic cloud. Instead, we define this feature to be magnetic ejecta and list the start and end times of the magnetic-flux-rope-like structure in the catalog where clearly present. Figure \ref{fig:exampleplots}a provides an example of an ICME with a clear flux rope structure observed by Juno during 2013 February (event 20130219). This event comprises a shock-like discontinuity (delineated by the fist solid vertical line), compressed sheath region, and a low-variance region of ejecta, which includes a magnetic flux rope structure (defined as between the dashed vertical lines).

Identification of potential ICME candidates becomes more difficult with increasing heliocentric distance. As ICMEs propagate away from the Sun, the likelihood of interaction with other ICMEs or solar wind transients increases. A previous study of ICMEs and SIRs identified at Ulysses whilst located at 5.3~AU found that 37\% of ICMEs had merged with SIRs \citep[][]{jian2008stream}. In this catalog, we have included ICMEs that have likely merged with such features, noting these as merged interaction regions (MIRs). An example of an MIR is presented in Figure \ref{fig:exampleplots}b (event 20130129). The event was observed by Juno during 2013 January, whilst the spacecraft was located at a heliocentric distance of $\sim$~2~AU. The MIR boundaries are delineated by the solid vertical lines and therefore include the compressed solar wind material of higher magnitude that follows the flux-rope-like structure.

In the identification of ICMEs observed by Juno, it is not possible to discern the separate sheath and magnetic ejecta regions in many of the events using only the magnetic field data available. Identifying the leading and trailing edges of magnetic ejecta can be highly subjective, especially with the lack of plasma data available. In this study, the strongest indication of ICME magnetic ejecta is the presence of magnetic-flux-rope-like structures associated with the ICMEs identified. We therefore include only the flux-rope-like structures associated with some events in the Juno catalog when analyzing trends related to ICME magnetic ejecta in this study (see Section \ref{sec:variation_distance}). 

\section{ICME Catalog} \label{sec:juno_catalog}

The ICME catalog (published in machine-readable format) lists the 80 events identified in the Juno cruise magnetometer dataset that meet the criteria detailed in Section \ref{sec:identification_criteria}. Table \ref{tab:database} provides a sample of the catalog. Events are listed chronologically and can be identified by their corresponding identification number (ID\#), which follows the date format `YYYYMMDD' at which the shock-like discontinuity associated with the event was first observed (the ICME start time). ICME end times are also provided, as well as the defined leading and trailing edges of flux-rope-like structures associated with the events where present. Where two or more flux ropes are associated with the same event, the ID\# is given a letter suffix (e.g. `a', `b', etc.). Thirty-two events have associated flux rope structures, two of which contain two flux-rope-like structures; a total of 34 flux ropes have therefore been identified in this study.

\begin{splitdeluxetable*}{lccccccllcBcllcclllh}
\tablecaption{Sample of Catalog Events Identified in the Juno Cruise Data Set.}
\label{tab:database}
\tablewidth{0pt}
\tablehead{
\colhead{Juno ID} & \multicolumn{3}{c}{HAE} & \multicolumn{3}{c}{HEE} & \colhead{ICME Start} & \colhead{ICME End} & \colhead{ICME B$_{mean}$} & \colhead{ICME B$_{max}$} & \colhead{FR Start} & \colhead{FR End} & \colhead{FR B$_{mean}$} & \colhead{FR B$_{max}$} & \colhead{FR type} & \colhead{Handedness} & \colhead{Quality} & \colhead{Notes} \\
\colhead{} & \colhead{r [AU]} & \colhead{lat $[^{\circ}]$} & \colhead{lon $[^{\circ}]$} & \colhead{r [AU]} & \colhead{lat $[^{\circ}]$} & \colhead{lon $[^{\circ}]$} & \colhead{} & \colhead{} & \colhead{[nT]} & \colhead{[nT]} & \colhead{} & \colhead{} & \colhead{[nT]} & \colhead{[nT]} & \colhead{} & \colhead{} & \colhead{} & \colhead{}
}
\decimals
\startdata
20110910 & 1.04 & -0.04 & 352.34 & 1.04 & -0.04 & 5.63 & 2011-09-09 18:01:30 & 2011-09-10 22:51:42 & 12.63 & 26.57 & \nodata & \nodata & \nodata & \nodata & \nodata & \nodata & 2 & ICME followed by HSS arriving 11 Sep (as shown by Wind plasma features). \\
20110917 & 1.07 & -0.05 & 0.22 & 1.07 & -0.05 & 6.19 & 2011-09-17 06:29:30 & 2011-09-18 21:57:31 & 9.67 & 14.99 & 2011-09-17 19:59:30 & 2011-09-18 13:17:30 & 11.84 & 14.99 & SEN & L & 1 & Clear shock, sheath, and ejecta with clear flux rope structure. Trailing edge of FR ambiguous due to discontinuity in N magnetic field component around 18 Sep 14:30. Possibly followed by flank of another ICME. \\
20110927 & 1.10 & -0.05 & 9.91 & 1.10 & -0.05 & 6.32 & 2011-09-27 01:02:30 & 2011-09-28 20:14:32 & 8.47 & 21.25 & \nodata & \nodata & \nodata & \nodata & \nodata & \nodata & 1 & Period of burst of activity, with middle structure clear ICME with strong shock and sheath structure preceding ejecta. \\
20111005 & 1.14 & -0.06 & 18.11 & 1.14 & -0.06 & 5.82 & 2011-10-05 21:28:31 & 2011-10-07 16:34:29 & 7.64 & 11.11 & 2011-10-06 07:18:30 & 2011-10-07 04:40:30 & 8.26 & 9.41 & \nodata & \nodata & 2 & ICME with a slightly rotating FR structure. More rotation is observed by Wind. \\
20111009 & 1.16 & -0.06 & 20.96 & 1.16 & -0.06 & 5.49 & 2011-10-09 02:46:30 & 2011-10-11 12:42:57 & 6.27 & 12.67 & 2011-10-10 04:06:30 & 2011-10-10 20:33:29 & 5.44 & 6.12 & ENW & L & 1 & Shock and sheath followed by FR-like structure with clear rotation of magnetic field in phi. \\
20111025 & 1.24 & -0.06 & 34.41 & 1.24 & -0.06 & 2.59 & 2011-10-25 14:22:41 & 2011-10-28 08:43:31 & 10.69 & 27.00 & 2011-10-26 00:40:31 & 2011-10-26 13:36:31 & 19.70 & 21.04 & SEN & L & 1 & Strong shock with broken sheath, followed by a clear FR-structure. Subject of \text{\citet{davies2020radial}}.\\
\enddata
\tablecomments{Each event number is identified by a unique Juno ID. The spacecraft heliocentric distance (r), latitude and longitude in both Heliocentric Aries Ecliptic (HAE) and Heliocentric Earth Ecliptic (HEE) coordinates are listed. ICME start and end times are given, along with the mean magnetic field (B$_{mean}$) maximum magnetic field (B$_{max}$) observed within the ICME. Likewise for each flux rope (FR) associated with an ICME. The type of flux rope is listed with the corresponding handedness being either left-handed (L) or right-handed (R). The final two columns list the quality of the event between 1 and 3, where 1 is high quality and 3 is the weakest, and notes on each event (hidden). Table \text{\ref{tab:database}} is published in its entirety in machine-readable format. A sample is presented here for guidance regarding its form and content.}
\end{splitdeluxetable*} 

The times given for each boundary are presented in  International Organization for Standardization (ISO) time format. In some cases, the trailing edge is rather ambiguous to define; in these cases, an alternative time has also been defined in parentheses. Events 20120512, 20120616, and 20131008 are listed despite significant data gaps obscuring the magnetic flux rope, where the true trailing edge and leading edge are not identifiable. Despite the large data gaps, the identification of events 20120512 and 20120616 was possible due to the clarity of the partial flux rope present, and all three events were observed by other spacecraft in conjunction with Juno. The missing data in these cases are indicated by an asterisk (\text{*}) following their ID\#. It is therefore only possible to include 32 of the 34 flux ropes identified in the analysis of the catalog. 

The catalog includes the heliocentric distance, latitude, and longitude at which the event was observed (calculated at the ICME start time) in both Heliocentric Aries Ecliptic (HAE) and Heliocentric Earth  Ecliptic (HEE) coordinates. Stating both coordinate systems allows the user to see the variation in event locations in a fixed system, and with respect to the Sun-Earth line for ease in identifying alignments with Earth and other spacecraft, respectively. The mean and maximum magnetic field strengths for each ICME and associated flux ropes are listed, calculated between the ICME start and end times, and the flux rope start and end times, respectively. Where flux ropes are present, we note the type of flux rope, based on the classifications of \citet{bothmer1998structure} and \citet{mulligan1998solar}, and the corresponding handedness, left (LH) or right (RH). 

The final two columns provide a description of each event and a subjective indication of the quality of the event between 1 and 3, where 1 indicates a high-quality event, and 3 indicates a poor-quality event. Considerations in assigning the quality of an event include the ease in defining the ICME start and end times (and flux rope boundaries where present), the overall magnetic field profile, magnetic field variance and smoothness of the flux rope rotation, and the presence/absence of data gaps: quality 1 events display a clear ICME structure (shock, sheath, and ejecta) with well-defined boundaries and no data gaps. Events with large data gaps leading to obscured ICME and flux rope boundaries are automatically assigned a quality of 3, whether or not they contain other high-quality features. 

\subsection{Summary of Catalog Events} \label{sec:catalog_summary}

\begin{figure}[t]
\centering
\includegraphics[width=\textwidth]{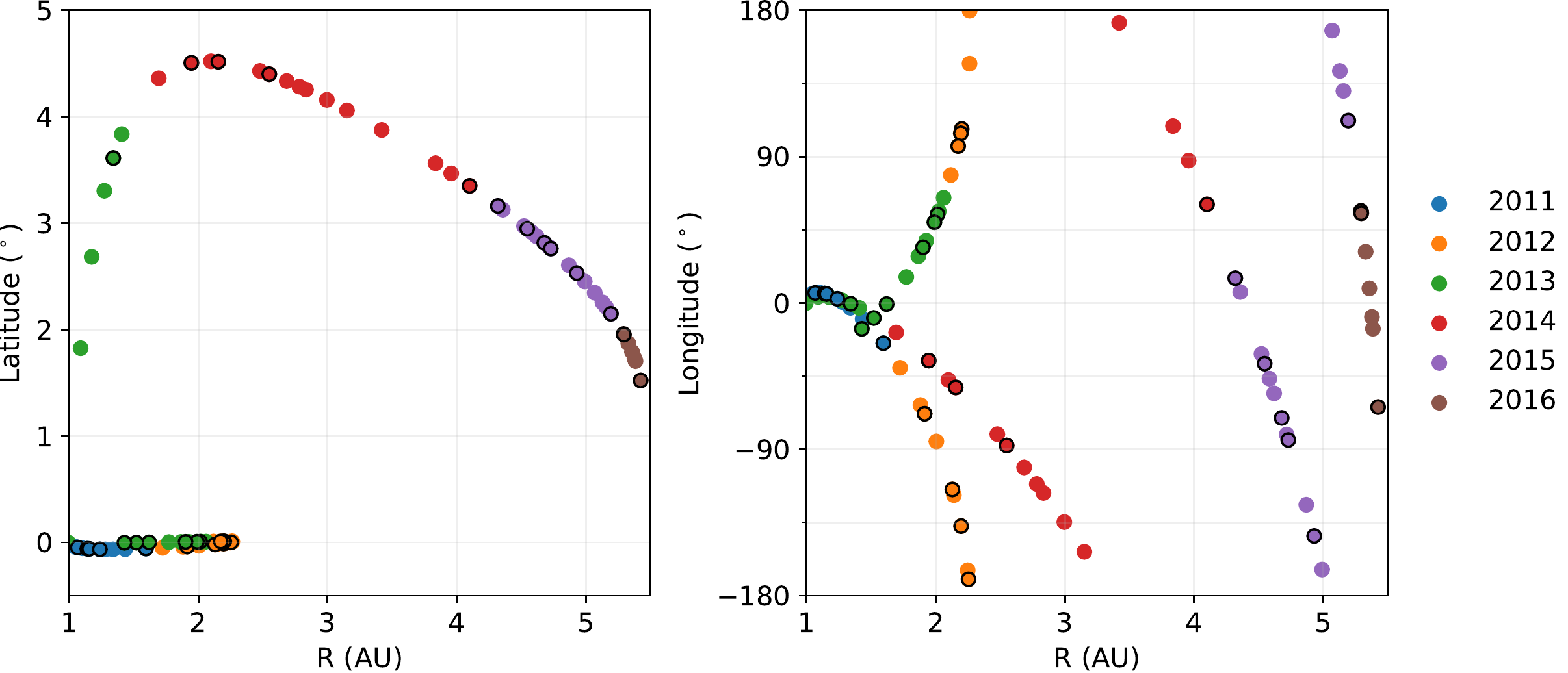}
\caption{Location of the Juno catalog events identified in this study in HEE coordinates. Latitude (left) and longitude (right) against heliocentric distance. Longitudes are given between 0$^\circ$ and $\pm$180$^\circ$, where 0$^\circ$ indicates the location of the Earth, 90$^\circ$ points west, and -90$^\circ$ points east. Events with associated flux ropes are highlighted by the black marker edges. The year of observation is indicated by the color key.}
\label{fig:event_distribution}
\end{figure}

Figure \ref{fig:event_distribution} shows the latitude (left) and longitude (right) against heliocentric distance at which the events in the catalog were observed in HEE coordinates. The colored markers correspond to the year in which they were observed. Markers with a black edge indicate that a flux rope structure was associated with the event. The positions are well distributed in both latitude and heliocentric distance over the cruise phase. Prior to the Earth gravity assist in 2013 October, event observations were measured very close to the ecliptic plane; 37 events, including 19 associated flux-rope-like structures, were observed within 0.2$^\circ$ of the ecliptic plane. The gravity assist increased the latitude of Juno to a maximum of 4.5$^\circ$ at 2.2~AU. Nine events and 4 associated magnetic-flux-rope-like structures are measured with latitudes between 1.8$^\circ$ and 4.5$^\circ$ following the assist. The 34 remaining events at heliocentric distances $>$2.5~AU decrease in latitude from 4.4$^\circ$ to 1.5$^\circ$. A previous study of events identified by Ulysses at latitudes less than 40$^\circ$ observed only small latitudinal variations in ICME properties \citep{ebert2009bulk}. Therefore, it is likely that the small range of latitudes sampled during the Juno cruise phase had little effect on the ICME properties measured in situ.

The longitudes are displayed between 0$^\circ$ and $\pm$180$^\circ$, where 0$^\circ$ indicates the location of the Earth. A total of 27 events and 10 flux rope structures were observed within 30$^\circ$ of the Earth and therefore could provide a basis for alignment studies with spacecraft located near the Earth. A previous such study involves the analysis of event 20111025, where ACE, Wind, and the ARTEMIS spacecraft were utilized to investigate the radial and longitudinal variation in ICME properties between the Earth and Juno at a distance of 1.24~AU \citep{davies2020radial}.

\begin{deluxetable*}{lcccccc|cllll}
\tablecaption{Yearly Summary of the ICME Catalog.}
\tablewidth{0pt}
\tablehead{
\multicolumn{6}{c}{\underline{ICMEs (Flux Ropes)}} & \multicolumn{1}{c|}{} & \multicolumn{1}{c}{} & \multicolumn{4}{c}{\underline{Flux Ropes Only}}\\
\colhead{Year} & \colhead{\#Events} & \colhead{\#Events Scaled} & \colhead{$\langle r_H \rangle$} & \colhead{$\langle B_{mean} \rangle$} & \colhead{$\langle B_{max} \rangle$} & \multicolumn{1}{c|}{} & \colhead{} & \colhead{\#LH} & \colhead{\#RH} & \colhead{\#HI} & \colhead{\#LI}
}
\decimals
\startdata
2011* & 10(5) & 28(14) & 1.24(1.24) & 7.36(5.69) & 14.68(11.09) & {} & {} & 4 & 0 & 2 & 2 \\
2012 & 15(8) & 17(9) & 2.11(2.12) & 3.87(3.82) & 7.65(5.17) & {} & {} & 1 & 5 & 2 & 3 \\
2013* & 17(7) & 29(12) & 1.65(1.69) & 4.49(4.58) & 8.24(5.52) & {} & {} & 4 & 3 & 4 & 3 \\
2014 & 15(5) & 17(6) & 2.85(2.54) & 2.77(3.52) & 5.56(4.35) & {} & {} & 2 & 2 & 1 & 3 \\
2015 & 16(6) & 16(6) & 4.78(4.73) & 1.55(1.50) & 2.59(1.78) & {} & {} & 3 & 3 & 1 & 5 \\
2016* & 7(3) & 14(6) & 5.35(5.34) & 1.63(2.03) & 2.80(2.27) & {} & {} & 1 & 2 & 1 & 2 \\
\hline
\textbf{Total} & 80(34) & {} & {} & {} & {} & {} & {} & 15 & 15 & 11 & 18 \\
\enddata
\label{tab:event_summary}
\tablecomments{Years with asterisks (\text{*}) indicate an incomplete dataset for that year. Number of events (\#Events), number of events scaled by the number of available days of data per year (\#Events Scaled), mean heliocentric distance ($\langle r_H \rangle$), average maximum magnetic field ($\langle B_{max} \rangle$) and mean magnetic field ($\langle B_{mean} \rangle$) are summarised. Numbers in brackets represent the same parameter calculated for only the flux ropes associated with the events in the catalogue. The number of left-handed (\#LH), right-handed (\#RH), high inclination (\#HI) and low inclination (\#LI) magnetic flux ropes are summarised in the right-hand side of the table.}
\end{deluxetable*}


Table \ref{tab:event_summary} provides a yearly summary of the events listed in the ICME catalog, where values in parentheses represent the same parameter for the flux ropes. Years that are incomplete or with significant data gaps are marked with asterisks (\text{*}). The number of events per year (\#Events) is given as well as the scaled number of events calculated using the number of days of data available for that year (\#Events Scaled), enabling a more consistent comparison across years of the cruise phase. There is a slight declining trend in the scaled number of events per year for both the ICME events and those with flux ropes associated; however, care must still be taken when interpreting this result due to the different data coverage across years and the increasing heliocentric distance at which events were observed. The scaled number of events per year is similar to the number of ICMEs listed in the \citet{richardson2010near} catalog identified at ACE (\url{http://www.srl.caltech.edu/ACE/ASC/DATA/level3/icmetable2.htm}), with the exceptions of 2012 and 2015, which are notably lower. This same pattern is observed if we compare the scaled number of flux ropes identified in this study with those noted as magnetic clouds at ACE. 

The mean heliocentric distance ($\langle r_H \rangle$) at which events were observed and the average maximum ($\langle B_{max} \rangle$) and mean magnetic field strengths ($\langle B_{mean} \rangle$) are also presented in Table \ref{tab:event_summary}. Comparing $\langle B_{max} \rangle$ and $\langle B_{mean} \rangle$ with $\langle r_H \rangle$, we see an overall decreasing trend each year with increasing heliocentric distance for both the ICMEs and the associated flux ropes. The relationship between magnetic field strength and heliocentric distance is explored further in Section \ref{sec:variation_distance}. 

\begin{figure}[t]
\centering
\includegraphics[width=\textwidth]{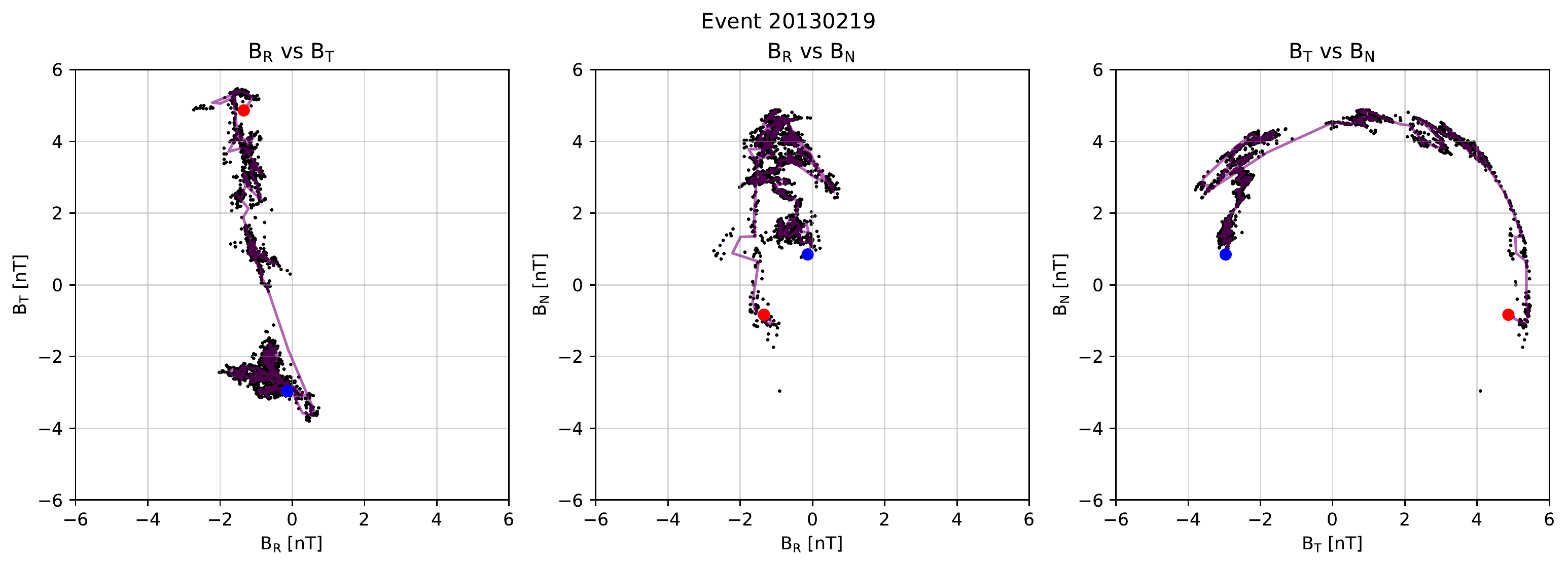}
\caption{Magnetic hodograms of the event 20130219 flux rope in RTN coordinates (magnetic time series presented in Figure \ref{fig:exampleplots}a). The three panels present different combinations of the magnetic field components, B$_\mathrm{R}$, B$_\mathrm{T}$, and B$_\mathrm{N}$. Start and end values of the magnetic field components are given by the red and blue markers, respectively. Hodograms give a sense of the rotation and handedness of the flux rope.}
\label{fig:hodogram_example}
\end{figure}

The final four columns of Table \ref{tab:event_summary} present a summary of the handedness (the number of left-handed and right-handed events per year) and inclinations (the number of high- and low-inclination events) of the flux ropes associated with the ICMEs in the catalog. Of the 32 complete flux ropes, it is possible using magnetic hodograms to determine the handedness of 30. Figure \ref{fig:hodogram_example} presents an example hodogram of the event 20130219 flux rope (the time series of which is presented in Figure \ref{fig:exampleplots}a). Most of the rotation is shown in the right-hand panel displaying the normal magnetic field component (B$_\mathrm{N}$) against the transverse component (B$_\mathrm{T}$), where the red marker indicates the start of the flux rope, and the blue marker the end. From this plot, we are able to deduce that the flux rope of event 20130219 is right-handed. 

The number of left-handed and right-handed magnetic flux ropes observed by Juno were equal: 15 left-handed and 15 right-handed. This result is to be expected as flux rope handedness can be associated with the solar hemisphere from which the CME (the solar-counterpart of the ICME) originated \citep{rust1994spawning}, however, the location of CME initiation has no such preference. The inclination of each flux rope is estimated by determining the flux rope type based on the classifications of \citet{bothmer1998structure} and \citet{mulligan1998solar}; high-inclination flux rope types are ESW, WNE, ENW, and WSE, and low-inclination flux rope types are SEN, NWS, SWN, and NES. We exclude event 20120728 as the classification is ambiguous, although certainly left-handed, leaving a total of 29 flux rope classifications. The total number of low-inclination flux ropes is greater than high-inclination flux ropes, in agreement with findings of previous studies \citep[e.g.][]{bothmer1998structure, good2016interplanetary}. As the heliocentric distance increases, the ratio of high-inclination to low-inclination flux rope types was found to decrease. Previous studies have found that ICMEs tend to deflect towards the solar equator, particularly during solar minimum when the solar wind is highly structured \citep{plunkett2001solar, cremades2006properties,wang2011statistical}. It has been suggested that ICME flux ropes may align with the HCS as they propagate \citep{yurchyshyn2008relationship, isavnin2014three}. \citet{isavnin2014three} found that although most of the deflection occurs below 30~R$_S$, a significant amount of deflection and rotation also occurs between 30~R$_S$ and 1~AU. The decrease in the ratio of high-inclination to low-inclination flux rope types may therefore be consistent with the suggestion that the magnetic flux rope axis of an ICME tends towards the local HCS (often close to the ecliptic plane during solar minimum) and therefore a lower inclination as it propagates. However, the Juno cruise phase took place during solar maximum where the HCS can be very highly inclined over a wide range of latitudes, with such a profile remaining relatively unchanged at larger heliocentric distances \citep{riley2002HCS}. A more detailed study of the inclination of the events and corresponding heliospheric environment is required to resolve whether events continue to rotate to lower inclinations or the local HCS beyond 1~AU, or whether the decrease in the ratio of high- to low-inclination flux rope types is an observational effect.

\subsection{Variation of Event Properties with Heliocentric Distance} \label{sec:variation_distance}

\begin{figure*}[t]
\centering
\includegraphics[width=\textwidth]{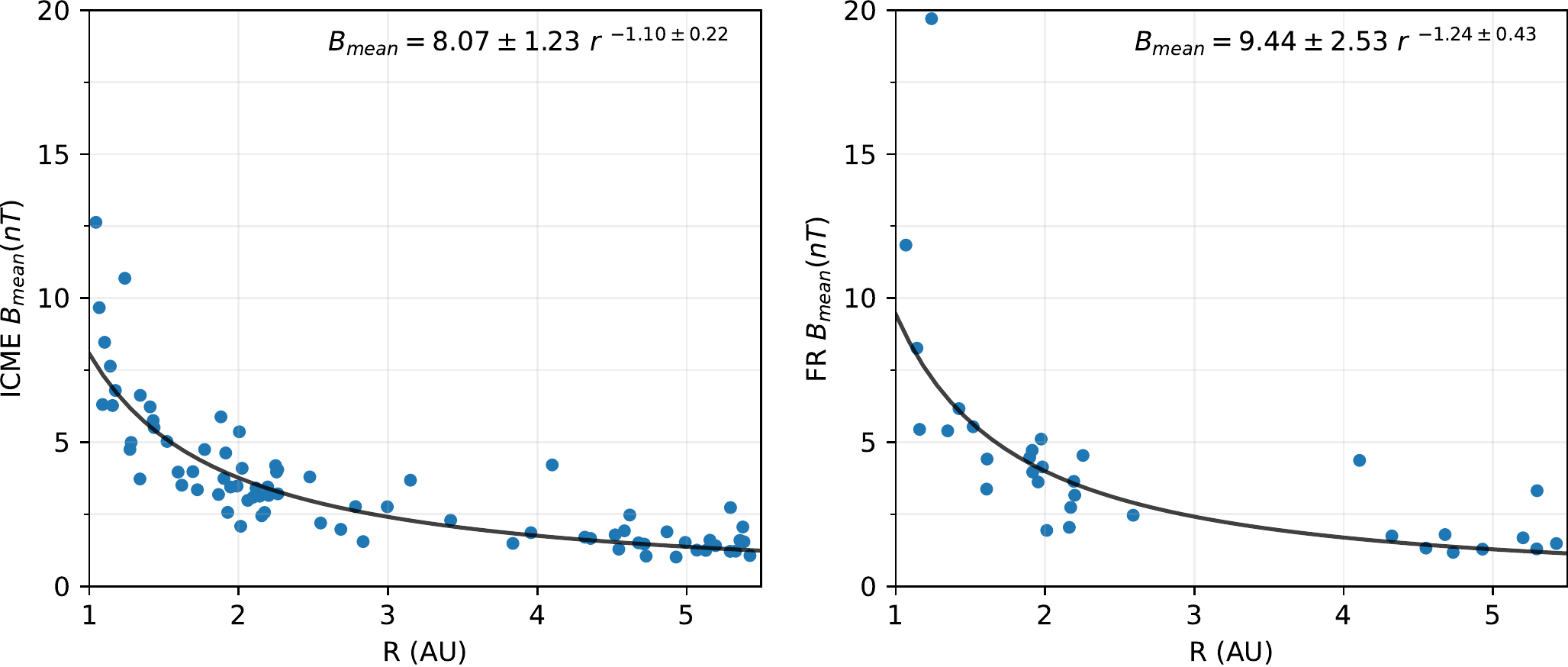}
\caption{Variation with heliocentric distance of the mean magnetic field strengths of the cataloged ICMEs (left) and flux-rope-like structures associated with the ICMEs (right).}
\label{fig:variation_B}
\end{figure*}

Assessment of how the magnetic field strength varies with heliocentric distance provides a measure of the global expansion of ICMEs. To investigate the relationship between magnetic field strength and increasing heliocentric distance, we have fitted the data using a least-squares fitting optimization model (scipy.optimize.least$\_$squares in Python) with a loss function of soft$\_$l1. This method provides a more robust fit to the data, where values further from the fit are given less weighting and therefore outliers have less influence on the final result. 

Figure \ref{fig:variation_B} shows the variation of the mean magnetic field strength of each cataloged ICME (left) and flux rope (right) with increasing heliocentric distance. The radial dependence of the mean ICME magnetic field decreases as $r^{-1.10 \pm 0.22}$ and the mean flux rope magnetic field decreases as $r^{-1.24 \pm 0.43}$. The variation of maximum field strength with increasing distance (not shown) was calculated for each ICME to decrease as $r^{-1.05 \pm 0.21}$ and each flux rope to decrease as $r^{-1.21 \pm 0.38}$. 
The relationship between the mean flux rope magnetic field strength and heliocentric distance is directly comparable to mean magnetic field relationships previously calculated using events identified by Ulysses. Between 1 and 5.4~AU, \citet{ebert2009bulk} found $B_{mean} \propto r^{-1.29 \pm 0.12}$, and similarly, \citet{richardson2014identification} found $B_{mean} \propto r^{-1.21 \pm 0.09}$. Both relationships are in strong agreement with the mean flux rope magnetic field relationship calculated in this study.

These relationships are very different to those found in the inner heliosphere ($<$1~AU). Previous studies that have used similar fitting to derive relationships of magnetic field strength with heliocentric distance $<$1~AU include \citet{gulisano2010global}, \citet{winslow2015interplanetary}, and \citet{salman2020radial}. \citet{gulisano2010global} used Helios 1 and 2 observations to obtain a mean magnetic field relationship of $B_{mean} \propto r^{-1.85 \pm 0.07}$. Similarly, \citet{winslow2015interplanetary} considered CMEs observed by MESSENGER and ACE to obtain a relationship of $B_{mean} \propto r^{-1.95 \pm 0.19}$. Considering 47 ICMEs observed in longitudinal conjunction by at least two of MESSENGER, Venus Express, STEREO, and ACE, \citet{salman2020radial} found the maximum magnetic field strength to vary as $B_{max} \propto r^{-1.91 \pm 0.25}$. These relationships suggest that the global expansion rate of ICMEs $<$~1~AU is faster than that of ICMEs $>$~1~AU.

\begin{figure*}[t]
\centering
\includegraphics[width=\textwidth]{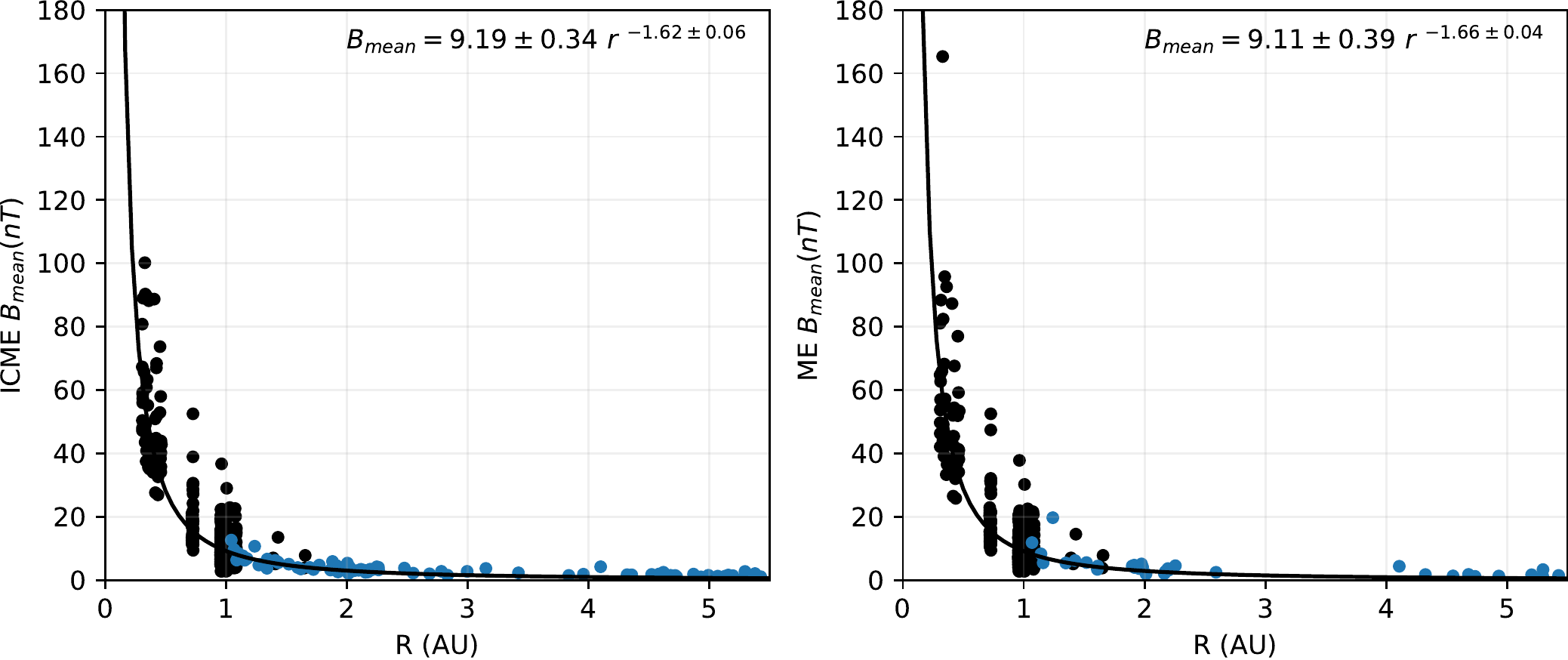}
\caption{Variation with heliocentric distance of the mean magnetic field strengths of events cataloged at Juno (blue) combined with those events listed in the HELCATS ICMECATv2.0 catalog (black) during the Juno cruise phase (from  2011 August to 2016 July). The combined data sets include ICMEs identified at MESSENGER, Venus Express, Wind, STEREO-A and B, and MAVEN to cover a heliocentric distance range of 0.3--5.4~AU. The relationships between the mean ICME magnetic field strength (left) and that of the magnetic ejecta (the combined magnetic ejecta entries listed in the HELCATS catalog and the magnetic flux ropes identified at Juno) with heliocentric distance (right) are presented.}
\label{fig:variation_HELCATS}
\end{figure*}

To extend the magnetic field relationships to cover a heliocentric distance range of 0.3--5.4~AU, we include ICME events listed in the HELCATS ICME catalog \citep[ICMECATv2.0;][]{moestl2017modeling,moestl2020prediction}. ICMECATv2.0 combines ICMEs identified in previous catalogs at MESSENGER \citep{winslow2015interplanetary}, Venus Express \citep{good2016interplanetary}, Wind \citep{nieves2018understanding}, and STEREO \citep{jian2018stereo}, with new entries identified by C. M\"ostl at those spacecraft and MAVEN, Ulysses, Solar Orbiter, Parker Solar Probe, and BepiColombo. The catalog currently lists 808 ICMEs observed between 2007 January and 2021 June \citep{moestl2020ICMECAT}. The distance at which events were observed is listed, as well as parameters such as the maximum and mean magnetic field strength. These parameters have been calculated over both the ICME as a whole and for the separate sheath and magnetic obstacle (MO) regions. Here, the definition of MO is equivalent to the more general magnetic ejecta term used throughout this study: the enhanced, low variance ICME material that typically follows the sheath. Thus, events listed in the ICMECATv2.0 include both flux-rope-like ejecta and more general magnetic ejecta. As described in Section \ref{sec:identification_criteria}, it has not been possible to discern the separate sheath and magnetic ejecta regions for many of the events identified in the Juno catalog, and therefore, we only use the flux-rope-like structures associated with some events as representative of magnetic ejecta at Juno. 

\begin{figure*}[t]
\centering
\includegraphics[width=\textwidth]{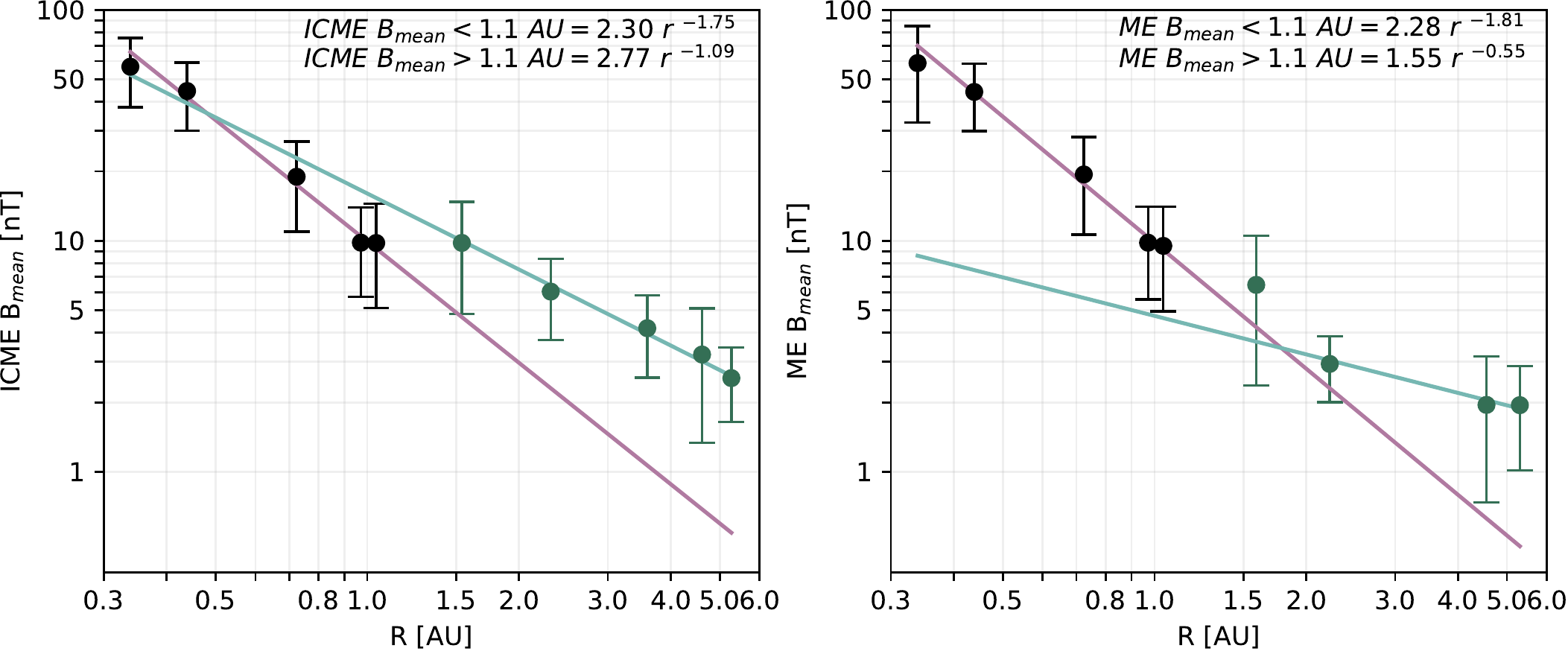}
\caption{Double-logarithmic plot of the observed mean magnetic field strength inside the ICMEs (left) and the magnetic ejecta (right) averaged in each radial bin, with increasing heliospheric distance. The error bars represent the standard deviation, $\sigma$, for values within each radial bin. The linear regression has been weighted using the inverse squared standard
deviation, $1/\sigma^2$, to produce two fits: averaged mean magnetic field values of radial bins less than 1.1~AU (purple) and greater than 1.1~AU (teal).}
\label{fig:log_linear_fit}
\end{figure*}

We filter the ICMECATv2.0 to include only entries recorded during the Juno cruise phase: from 2011 August to 2016 July. The remaining 425 ICME entries observed by MESSENGER, Venus Express, STEREO-A, STEREO-B, Wind, and MAVEN have been combined with the Juno catalog entries to produce an ICME sample size of 502 and a magnetic ejecta sample size of 452. Figure \ref{fig:variation_HELCATS} presents the mean magnetic field strengths of each cataloged ICME (left) and the HELCATS magnetic ejecta entries combined with the flux ropes identified at Juno (right) as a function of heliocentric distance. The robust power-law fit to the mean magnetic field data results in $B_{mean} \propto r^{-1.62 \pm 0.06}$ for ICMEs and $B_{mean} \propto r^{-1.66 \pm 0.04}$ for the combined HELCATS magnetic ejecta and Juno flux ropes. In both the ICME and magnetic ejecta fits, events observed at heliocentric distances $>$3~AU lie above the best-fit line, suggesting the events at lesser heliocentric distances had a greater influence on the overall fitting result. This is likely due to the greater number of events in the HELCATS catalog in comparison to the Juno catalog, and the greater variability in magnetic field strengths at lesser heliocentric distances. The difference in global expansion rate at different heliocentric distances is explored further in Figure \ref{fig:log_linear_fit}. 

The combined magnetic ejecta relationship can be compared to those previously determined by combining the Ulysses data with other spacecraft datasets: \citet{richardson2014identification} included the 103 ICMEs observed by the Helios 1 and 2 spacecraft and those identified in the \citet{richardson2010near} catalog, to calculate a relationship of $B \propto r^{-1.38\pm0.03}$ between 0.3 and 5.4~AU. Similarly, \citet{liu2005statistical} found that $B \propto r^{-1.40\pm0.08}$ by including ICMEs identified at Helios, Wind, and ACE, and \citet{wang2005characteristics} found that $B \propto r^{-1.52}$ by including ICMEs identified at Helios, the Pioneer Venus Orbiter, and ACE. These relationships result in higher powers than those calculated in this study; however, the magnetic field strength relationships calculated over heliocentric distances between 0.3 and 5.4~AU are mostly in agreement with power dependencies between those calculated over heliocentric distances below 1~AU and those beyond 1~AU. It is possible that the difference in powers between studies may due to the weaker Solar Cycle 24 in which the Juno cruise phase took place in comparison to previous solar cycles, as although ICME rates have been found to be consistent between solar cycles, the lower external pressure allows for more rapid ICME expansion \citep[e.g.][]{gopalswamy2014anomalous}.

Figure \ref{fig:log_linear_fit} presents the variation of the mean ICME and magnetic ejecta field strengths of the combined catalogs with heliocentric distance on a double-logarithmic plot. The data have been sorted into radial bins of size 0.1~AU below 1.1~AU (black), and $\sim$~1~AU above 1.1~AU (green). The differences in radial bin size were chosen to even out the number of events in each bin, with the exceptions of 0.9--1~AU which includes the many events observed by Wind and STEREO-A, and 1--1.1~AU which includes mostly events observed by STEREO-B. The fitting has been performed using a linear regression model (sklearn.linear$\_$model.LinearRegression of the scikit-learn package in Python) weighted using the inverse squared standard deviation, $1/\sigma^2$. The fitting produces relationships of $B_{mean} \propto r^{-1.75}$ and $B_{mean} \propto r^{-1.09}$ for ICMEs observed at heliocentric distances below (purple) and above 1.1~AU (teal), respectively. A similar difference between fits is found using the magnetic ejecta values: $B_{mean} \propto r^{-1.81}$ and $B_{mean} \propto r^{-0.55}$. Qualitatively, the calculated relationships fit well to the data in each case with the exception of the magnetic ejecta relationship $>$~1.1~AU, where the latter three bins are well fit but not the first. A qualitatively better fit is achieved by removing the weighting (not shown) to produce a relationship of $B_{mean} \propto r^{-0.91}$. The fitting demonstrates that there is a clear difference in expansion rate for ICMEs observed at lesser heliocentric distances and those observed farther out. This result is consistent with the findings of previous studies \citep[e.g.][]{lugaz2020inconsistencies}, where expansion in the innermost heliosphere is dominated by the large total pressure inside the ICME, resulting in a more rapid decrease than beyond 1~AU, where expansion is dominated by the decrease of the IMF ($\sim r^{-1}$). The powers calculated for magnetic ejecta beyond 1.1~AU indicate a lower expansion rate than that of the IMF, suggesting that perhaps MIRs and other interactions become more dominant in the evolution of the ME magnetic field at these larger heliocentric distances.

\section{Summary} \label{sec:summary}

We have identified 80 ICME candidates, 32 of which have associated flux-rope-like structures, in the magnetometer data taken during the Juno mission cruise phase. Two of the events each had two flux-rope-like structures associated with them, and therefore a total of 34 flux ropes have been identified in this study. Each event meets the criteria of an enhanced magnetic field magnitude at least twice that of the expected ambient IMF, a magnetic field profile typical of an ICME e.g. a shock, sheath, and region of magnetic ejecta, and a duration on the order of at least one day. The catalog of events covers a heliocentric distance range of 1--5.4 AU, the first opportunity to study ICMEs beyond 1~AU since the Ulysses mission. 

For each event, the mean and maximum magnetic field strengths of each ICME have been recorded. We found that:

\begin{itemize}
    \item The number of events per day of available data per year showed a slightly decreasing trend over the cruise phase. 
    \item The mean ICME magnetic field decreases as $r^{-1.10 \pm 0.22}$ and the maximum magnetic field decreases as $r^{-1.05 \pm 0.21}$ with increasing heliocentric distance.
\end{itemize}

For the 34 flux-rope-like structures associated with the ICME candidates, the flux rope type, handedness, and magnetic field strengths (maximum and mean) within the flux rope were recorded where possible. We found that: 

\begin{itemize}
\itemsep-0.5em
    \item The number of left-handed and right-handed magnetic flux ropes observed by Juno was equal:  15 left-handed and 15 right-handed.
    \item As the heliocentric distance increases, the ratio of high-inclination to low-inclination flux rope types tends to decrease. This result is consistent with previous studies that suggest the magnetic flux rope axis of an ICME tends towards the ecliptic plane or local HCS as it propagates, but could also simply be a consequence of spacecraft location or path taken through the ICME. 
    \item The mean flux rope magnetic field decreases as $r^{-1.24 \pm 0.43}$ and the maximum magnetic field decreases as $r^{-1.21 \pm 0.38}$ with increasing heliocentric distance. The mean magnetic flux rope relationship is directly comparable to those derived using events identified at Ulysses and found to be in strong agreement.
\end{itemize}

Combining the Juno catalog with the HELCATS catalog, we created a dataset covering 0.3--5.4~AU. We found: 
\begin{itemize}
\itemsep-0.5em
    \item A robust power-law fit to the mean magnetic field data results in $B_{mean} \propto r^{-1.62 \pm 0.06}$ for ICMEs and $B_{mean} \propto r^{-1.66 \pm 0.04}$ for the combined HELCATS magnetic ejecta and Juno flux ropes. The power of these relationships is lower than that of similar relationships found previously by combining Ulysses ICMEs with other data sets, likely due to the difference in the number of events between the HELCATS and Juno catalog, and therefore the greater influence events at lesser distances had on the nonlinear fitting.
    \item By sorting the combined HELCATS and Juno catalogs into bins of radial heliocentric distance, we show on a double-logarithmic plot the clear difference in global expansion rates between events observed at lesser heliocentric distances and those observed farther out in the heliosphere. This result is consistent with the findings of previous studies where expansion in the innermost heliosphere is more rapid than that beyond 1~AU where it is dominated by the decrease of the IMF. The fitting of the magnetic ejecta data beyond 1.1~AU suggests a lower expansion rate than that of the IMF, perhaps an indication that MIRs and other interactions become more dominant in the evolution of the ME magnetic field at these larger heliocentric distances.
\end{itemize}

The catalog of events at Juno provides a basis for further studies of ICME properties beyond 1~AU and conjunctions between Juno and other spacecraft could be explored to further understanding of ICME evolution over larger heliocentric distances. Analysis of ICMEs observed by Juno has contributed to our understanding of ICME evolution beyond 1~AU and forms a good basis on which other future missions to the outer planets, such as JUICE, may build upon.

\begin{acknowledgments}
We have benefited from the availability of the Juno cruise phase data and thus would like to thank the Juno MAG Principal Investigator, J.E.P. Connerney, and the instrument team. We also thank the PDS (\url{https://pds.nasa.gov}) archive for its distribution of data. The specific Juno MAG and JADE cruise data sets are available on PPI/PDS: \citet{pdsJUNOmag, pdsJUNOjade}. This research was supported by funding from the UKRI Science and Technology Facilities Council studentship ST/N504336/1 (E.E.D) and by NASA grants 80NSSC19K0914 (R.M.W. and E.E.D) and 80NSSC20K0700 (N.L.). C. M. thanks the Austrian Science Fund (FWF): P31659-N27, P31521-N27.
\end{acknowledgments}

\vspace{5mm}
\software{spiceypy \citep{annex2020spiceypy}, scipy \citep{virtanen2020scipy}, scikit-learn \citep{scikit-learn}}


\bibliography{bibliography}{}
\bibliographystyle{aasjournal}


\end{document}